\documentclass[aps,preprint,amsmath,amssymb,amsfonts,nofootinbib]{revtex4}
\usepackage{epsfig}
\usepackage{graphicx}
\usepackage{dcolumn}
\usepackage{bm}
\usepackage{amsthm}
\usepackage{amsmath}
\usepackage{color}

\newcommand{\beq}{\begin{equation}}
\newcommand{\eeq}{\end{equation}}
\newcommand{\bea}{\begin{eqnarray}}
\newcommand{\eea}{\end{eqnarray}}

\begin{document}
\title{Horava-Lifshitz Black Hole Hydrodynamics}
\author{Christopher Eling$^1$}
\author{Yaron Oz$^2$}
\affiliation{$^1$ Max Planck Institute for Gravitational Physics, Albert Einstein Institute, Potsdam 14476, Germany}
\affiliation{$^2$ Raymond and Beverly Sackler School of Physics and Astronomy, Tel Aviv University, Tel Aviv 69978, Israel}

\date{\today}
\begin{abstract}
We consider the holographic hydrodynamics of black holes in generally covariant gravity theories with
a preferred time foliation. Gravitational perturbations in these theories have spin two and spin zero helicity modes
with generically different speeds.
The black hole solutions possess a spacelike causal boundary called the universal horizon. We relate
the flux of the spin zero perturbation across the universal horizon
to the new dissipative transport in Lifshitz field theory hydrodynamics found in  arXiv:1304.7481. We construct in detail the hydrodynamics of one such black hole solution, and calculate the ratio of the shear viscosity to the entropy density.

\end{abstract}

\pacs{...}

\maketitle

\section{Introduction}

A generic, generally covariant model of local Lorentz violating gravity is Einstein-aether theory \cite{AEtheory}.
In this theory the symmetry is broken by the aether covector $v_A$, which is a dynamical field that is constrained to be unit timelike.
As a consequence, the theory has in general spin-2, spin-1, and spin-0 gravitational wave polarizations traveling at different speeds.
A particular choice for the aether field is to be hypersurface orthogonal, thus determining a preferred time foliation of space-time.  In this case the Einstein-aether theory can be reduced \cite{Jacobson:2010mx, Blas:2010hb} to the Horava-Lifshitz theory \cite{Horava:2009uw}.

Spherically symmetric asymptotically flat black hole solutions were first constructed in \cite{Eling:2006ec,Barausse:2011pu}.
The absence of local Lorentz symmetry has profound effects on black hole thermodynamics.
Causality is no longer determined by the light cone, and
the presence of multiple horizons seems to conflict with both the Zeroth and Second Laws  \cite{Dubovsky:2006vk,Eling:2007qd,Jacobson:2008yc}.
Still, one can argue (see, e.g. \cite{Barausse:2011pu,Blas:2011ni}) that there is a natural notion of causality defined by the preferred time foliation itself.  At spatial infinity the time translation Killing vector and the aether are naturally aligned. However deep in the bulk,  surfaces of constant preferred time bend and these two vectors eventually become orthogonal on a spacelike hypersurface. This is equivalent to the statement that the aether time $\tau \rightarrow \infty$ on this hypersurface, called  ``universal horizon". Beyond the universal horizon even instantaneously propagating modes are causally disconnected from spatial infinity.
Studies of the universal horizon suggest  an associated temperature, and a first law of thermodynamics  \cite{Berglund:2012bu,Berglund:2012fk,Mohd:2013zca,Cropp:2013sea}.

In this paper we will be interested in studying the Einstein-aether/Horava-Lifshitz theory with a negative cosmological constant,
in the context of holography and the equivalence to a non-gravitational field theory.
Holography is expected to relate these gravitational systems to field theories with broken Lorentz invariance
in one lower space dimension. When the bulk aether field is hypersurface orthogonal, it
induces a foliation at the boundary.
The corresponding boundary field theories are known
as Lifshitz field theories. In such field theories  Lorentz invariance is broken.

Lifshitz field theories exhibit an anisotropic scaling of space and time (Lifshitz scaling) $x^i\rightarrow \lambda x^i, i=1,...,d, t\rightarrow \lambda^z t$.
$z$ is called a dynamical exponent and is a measure of the anisotropy. In relativistic conformal field theories (CFTs) $z=1$.  The dynamical exponent differs from one in general Lifshitz theories. Examples of Lifshitz dual theories in $2+1$ dimensional space-time
are given by quantum critical points (QCPs). Such theories describe phase transitions at zero temperature driven by quantum fluctuations
\cite{subir}.

At zero temperature the correlation length $\xi$ at the QCP diverges.
Raising the temperature, one finds a quantum critical regime, where the system properties are constrained
by the anisotropic scaling at the QCP.
Denote by $L$ a characteristic  length scale of the system and by $T$ the temperature. Hydrodynamics provides
a good description in the quantum critical regime at the range
of scales $\xi \gg L \gg l_T \sim \frac{1}{T^{\frac{1}{z}}}$. The hydrodynamics expansion parameter is
the dimensionless ratio $\frac{l_T}{L}$.

Since boost invariance is broken in Lifshitz field theories, the stress-energy tensor
is no longer symmetric. The asymmetric term is associated with the foliation 1-form.
With rotation invariance, the hydrodynamics of Lifshitz field theories exhibits one new dissipative transport coefficient
at the first dissipative order
found in \cite{Hoyos:2013eza,Hoyos:2013qna}.

The gauge/gravity duality relates field theories at finite temperature to black holes in one higher space dimension. We will be interested in the out of equilibrium dynamics of such black holes. We will work in the hydrodynamic regime, which is described by black hole deformations, order by order in a derivative
expansion. The field theory Navier-Stokes equations
are the gravity constraint equations \cite{Bhattacharyya:2008jc}, which are also the horizon evolution equations \cite{Eling:2009sj}.

Gravity is non-dissipative, however the horizon being a one way membrane introduces an effective dissipation: what goes
in cannot go out. We propose that this boundary condition should be imposed at the universal horizon in these theories. Gravitational
backgrounds with preferred foliation allow the propagation of spin-2 and spin-0 helicity gravitons.
In thermal field theory language these two modes correspond to two possible channels of dissipation. The
dissipation associated with the spin-2 helicity mode is seen in the standard relativistic hydrodynamics as the viscosity
terms in the symmetric stress-energy tensor. The dissipation associated with the spin-0 helicity should be related to a new transport coefficient.
We will argue that this is the new dissipation in Lifshitz field theory hydrodynamics appearing in the asymmetric part of the stress-energy tensor, and
discovered in
 \cite{Hoyos:2013eza,Hoyos:2013qna}.

  In the special case where $z=1$, an analytic
   asymptotically Lifshitz plane symmetric black hole solution to Horava-Lifshitz gravity
   is available  \cite{Griffin:2012qx,Janiszewski:2014iaa}.
   We will show that this new transport coefficient is zero is this case, and we will calculate the shear viscosity ratio $\eta/s$.

The paper is organized as follows.
In section two we will discuss Einstein-aether and Horava-Lifshitz theories, the action and the classical field
equations, and asymptotically Lifshitz solutions  when a negative cosmological constant is added.
In section three we will discuss Holographic Lifshitz Hydrodynamics.
We will briefly review Lifshitz field theory hydrodynamics and construct
the boundary stress-energy tensor. We will explain how its asymmetric part arises and exhibit the new dissipative
transport coefficient associated with it.
We will present the constraint equations and in particular the null focusing equation that is the gravitational
counterpart of the entropy current divergence in field theory hydrodynamics. We will show
where the new channel for dissipation comes from and connect it to the flux of the spin-0 helicity mode through the universal horizon.
In section four we will analyze in detail the hydrodynamics at the first derivative
order of a particular $z=1$ analytic solution of Horava-Lifshitz black brane hydrodynamics. We will show that
the new dissipative transport associated with the lack of Lorentz invariance vanishes in this case. We will calculate
the ratio of the shear viscosity
to the entropy density and show that it deviates from the general relativity result.
Section five is devoted to a discussion of open problems.

\section{Einstein-aether and Horava-Lifshitz}

\subsection{Einstein-aether action and field equations}

In the following we will work in four-dimensional  space-time dimensions (the generalization to other dimensions is
straightforward).
The action for Einstein-aether theory is given by
\begin{align}
S_{ae} = \frac{1}{16\pi G_{ae}} \int d^4 x \sqrt{-g} L_{ae} \label{totalaction} \ ,
\end{align}
where $L_{ae} = R + L_{vec}$ \ ,
\begin{align}
-L_{vec} = K^{AB}{}_{CD} \nabla_A v^C \nabla_B v^D - \lambda(v^2+1) \ ,
\end{align}
with ``kinetic" tensor defined as
\begin{align}
K^{AB}{}_{CD} = c_1 g^{AB} g_{CD} + c_2 \delta^A_C \delta^B_D + c_3 \delta^A_D \delta^B_C - c_4 v^A v^B g_{CD} \ .
\end{align}
This is the most general effective action for a timelike unit vector field at 2nd order in derivatives.

Varying this action with respect to the metric, vector field, and the Lagrange multiplier $\lambda$, one finds the following field equations
\begin{align}
G_{AB} = T^{ae}_{AB}, ~~ E_A = 0, ~~ v^2 = -1 \ .
\end{align}
The aether stress tensor is given by
\begin{align}
T^{ae}_{AB} = \lambda v_A v_B + c_4 a^{(v)}_A a^{(v)}_B - \frac{1}{2} g_{AB} Y^C{}_D \nabla_C v^D + \nabla_C X^C{}_{AB} + c_1[(\nabla_A v_C)(\nabla_B v^C)-(\nabla^C v_A)(\nabla_C v_B)] \ ,
\end{align}
where
\begin{align}
Y^A{}_B =& K^{AC}{}_{BD} \nabla_C v^D  \ , \\
X^C{}_{AB} =& Y^C{}_{(A} v_{B)} - v_{(A} Y_{B)}{}^C + v^C Y_{(AB)} \ ,  \
\end{align}
and $a^{(v)}_A = v^B \nabla_B v_A$ is the aether acceleration (which we distinguish from the fluid acceleration defined in the next section). The aether field equation is
\begin{align}
E_A = \nabla_B Y^B{}_A + \lambda v_A + c_4 (\nabla_A v^B) a^{(v)}_B \label{aefieldeqn} \ .
\end{align}

\subsection{Hypersurface orthogonality and  Horava-Lifshitz gravity}

Consider now the case where the aether field is hypersurface orthogonal. This means that the twist vanishes
\begin{align}
v_{[A} \nabla_B v_{C]} = 0 \ .
\end{align}
Since the squared twist also vanishes
\begin{align}
\omega^2 = (\nabla_A v_B)(\nabla^A v^B)-(\nabla_A v_B)(\nabla^B v^A)+a^2 \ ,
\end{align}
adding a twist squared term to the action doesn't affect the solutions.
We can therefore eliminate either $c_1$, $c_3$ or $c_4$ in the action. Here we will choose to eliminate $c_1$, i.e we take $c_1=0$ from now on.

Hypersurface orthogonality implies the co-vector is the gradient of a scalar
\begin{align}
v_A = \frac{-\partial_A \phi}{\sqrt{g^{CD} \partial_C  \phi \partial_D \phi}} \ .
\end{align}
One can show that hypersurface orthogonal solutions of Einstein-aether theory are also solutions to Horava-Lifshitz gravity \cite{Jacobson:2010mx}. The connection can be made explicit by choosing coordinates such that $\phi=\tau$, where $\tau$ is the preferred foliation of time. In this gauge the Einstein-aether action reduces to the generic 3+1 form of the Horava-Lifshitz action (e.g. \cite{Griffin:2012qx})
\begin{align}
S_{HL} = \frac{1}{16\pi G_H} \int d\tau d^3 x \sqrt{\gamma} \left(K_{ab} K^{ab} - (1-\lambda) K^2 +(1+\beta) R^{(3)} + \tilde{\alpha} \frac{\nabla_a N \nabla^a N}{N^2} \right)
\ \label{HLaction}.
\end{align}
Here $K_{ab}$ is the extrinsic curvature of the preferred time slices, $\gamma_{ab}$ the spatial metric on the slices, $R^{(3)}$ the intrinsic Ricci scalar and $N$ is the lapse function, i.e. $v_A = -N \delta^\tau_A$. The mapping between the constants is given by
\begin{align}
1+\lambda = \frac{1+c_2}{1-c_3}, ~~ \tilde{\alpha} = \frac{c_4}{1-c_3}, ~~ \frac{G_H}{G_{ae}} = 1+\beta = \frac{1}{1-c_3} \ .
\end{align}
In generic Einstein-aether theory there are five propagating degrees of freedom with spin-2, spin-1, and spin-0 helicities \cite{Jacobson:2004ts}. In Horava-Lifshitz the spin-1 mode is non-propagating. The squared speeds of the remaining modes are given (in four dimensions) by \cite{Griffin:2012qx}
\begin{align}
s_2^2 = \frac{1}{1-c_3}, ~~ s_0^2 = \frac{(c_2+c_3)(3-c_4)}{c_4(1-c_3)(3-4c_2+c_3)} \label{speeds} \ .
\end{align}

\subsection{Asymptotically Lifshitz solutions}

In \cite{Janiszewski:2014iaa} asymptotically Lifshitz and AdS solutions were studied in detail. For additional studies of these solutions, see \cite{Janiszewski:2014ewa, solutions,Sotiriou:2014gna}. In this case one adds a negative cosmological constant $\Lambda$ to the action above
\begin{align}
L_{total} = \frac{1}{16\pi G_{ae}} \left(\Lambda + L_{ae} \right) \ .
\end{align}
We consider a metric and aether ansatz of the form
\begin{align}
ds^2 = F(\rho) dt^2 - 2 G(\rho) dt d\rho + \rho^2 dx_i dx^i, \\
v_A dx^A = \frac{G(\rho)^2-F(\rho)K(\rho)^2}{2 K(\rho) G(\rho)} dt + K(\rho) d\rho \ .
\label{ansatz}
\end{align}
Foreshadowing the holographic setup, we take $x^A = (x^\mu, \rho)$ with field theory coordinates $x^\mu = (t,x^i)$.
The Lifshitz scaling reads $\rho \rightarrow \lambda^{-1} \rho, x^i\rightarrow \lambda x^i, t\rightarrow \lambda^z t$.

Near infinity, the solution is required to approach
\begin{align}
ds^2 \sim - \rho^{2z} dt^2 + 2 \rho^{z-1} dt d\rho + \rho^2 dx_i dx^i \\
K(\rho) \sim \frac{1}{\rho} \ ,
\end{align}
It turns out that consistency of the field equations with this ansatz requires
\begin{align}
c_4 = \frac{z-1}{z}, \Lambda = - \frac{(1+z)(2+z)}{2} \ .
\end{align}
Generically, solutions with $z \neq 1$ can only be found numerically. In the case where $z=1$ and $c_4=0$ (asymptotically AdS) an analytic solution was found in \cite{Janiszewski:2014iaa}. We will consider this case later in the paper. We expect that long wavelength, long time perturbations of these Lifshitz black brane solutions to be described by the hydrodynamics of Lifshitz field theories, which we describe in the following section.

\section{Holographic Lifshitz Hydrodynamics}

\subsection{Lifshitz field theory hydrodynamics}

Since boost invariance is explicitly broken in Lifshitz field theories, the conserved stress-energy tensor
is not necessarily symmetric. In order to see its  asymmetric part, we have to construct it not as a response of the action $S$
to a change in a background metric $h_{\mu\nu}$, but rather as a response to a change in the vielbein $e^{\mu}_a$ (by $a$ we denote tangent space indices)
\begin{equation}
T^a_{\mu} = -\frac{1}{e}\frac{\delta S}{\delta e^{\mu}_a} \ .
\label{Te}
\end{equation}
The vielbein encodes both the metric data $h_{\mu\nu} = e^a_{\mu}e_{\nu}^b \eta_{ab}$, and
the foliation data $v_{\mu} = e_{\mu}^a v_{a}$, where $v_a=(1,0..,0)$.

Using (\ref{Te}) one has
\begin{equation}
T_{\mu\nu} = \Theta_{\mu\nu} + J_{\mu}v_{\nu} \ ,
\label{stress}
\end{equation}
where
\begin{equation}
\Theta_{\mu \nu} = \frac{2}{\sqrt{-h}}\frac{\delta S}{\delta h^{\mu \nu}},~~~~~
J_{\mu} = \frac{1}{\sqrt{-h}} \frac{\delta S}{\delta v^\mu} \ . \label{TJ}
\end{equation}
We see from (\ref{stress}) that
the asymmetric part of the stress-energy tensor arises from $J_{\mu}v_{\nu}$ and is directly connected to the foliation data.

Consider next the hydrodynamics of Lifshitz field theories.
The hydrodynamic stress-energy tensor in the uncharged case is  expressed in terms of the temperature $T$, the velocity vector $u^{\mu}$
normalized as $u_{\mu}u^{\mu}=-1$,
and their derivatives via the constitutive relations.
The hydrodynamics equations are the conservation law of the stress-energy tensor
$\partial_{\mu}T^{\mu\nu}=0$.
As above, since boost invariance is explicitly broken, the stress-energy tensor can have an asymmetric part.
 Assuming rotation invariance, the asymmetric
term shows up at the first viscous order  \cite{Hoyos:2013eza,Hoyos:2013qna}.

The energy-momentum tensor in the Landau frame $T^{\mu\nu} u_\nu=-\varepsilon u^\mu$ takes the form
\begin{equation}
T^{\mu\nu}=\varepsilon u^\mu u^\nu +p P^{\mu\nu}+\pi_S^{(\mu\nu)}+\pi_A^{[\mu\nu]} +(u^\mu\pi_A^{[\nu\sigma]}+u^\nu\pi_A^{[\mu\sigma]})u_\sigma ,
\end{equation}
with $\pi^{(\mu\nu)}_S u_\nu=0$.
At first order in derivatives $\pi_S^{(\mu\nu)}$ includes the shear and bulk viscosities.
The antisymmetric part of the stress-energy tensor reads at first order
\begin{equation}
\pi^{[\mu\nu]}_A = -\alpha u^{[\mu}a^{\nu]} \ ,
\end{equation}
where $a^{\mu} = u^{\nu}\partial_{\nu}u^{\mu}$ is the fluid acceleration, and $\alpha$ is a dissipative transport coefficient.
It contributes to the divergence of the entropy current $s^{\mu} = s u^{\mu}$
\begin{equation}
\partial_{\mu}s^{\mu}  = \frac{2\eta}{T} \sigma_{\mu \nu} \sigma^{\mu \nu} + \frac{\zeta}{T} (\partial_\mu u^\mu)^2 + \frac{\alpha}{T} a_{\mu}a^{\mu} \label{hydroentropylaw}\ .
\end{equation}
Here $\sigma_{\mu \nu} = P^{\lambda}_{\mu} P^{\sigma}_{\nu} \partial_{(\lambda} u_{\sigma)} - \frac{1}{3} P_{\mu \nu} \partial_\lambda u^\lambda$ is the fluid shear tensor, with $P_{\mu \nu} = h_{\mu \nu}+u_\mu u_\nu$ the projection tensor orthogonal to $u^\mu$.
The local form of the second law of thermodynamics $\partial_{\mu}s^{\mu} \geq 0$ requires that
$\alpha \geq 0$, in addition to the usual positivity conditions on the shear  and bulk viscosities $\eta$ and $\zeta$, respectively.
We will argue that the $a_{\mu}a^{\mu}$ entropy production term corresponds to the flux of spin-0 helicity graviton through the universal horizon.

\subsection{The boundary stress-energy tensor}
\label{boundarysection}

In the following we derive the boundary stress-energy tensor from
the gravity side. Suppose that we have the on-shell classical action $S_{cl}$, which is a function of boundary data $h_{\mu \nu}$ and $v_\mu$. This classical action is invariant under diffeomorphisms in the boundary generated by $\xi^\mu$. One finds
\begin{align}
\delta_\xi S_{cl} = \int \left(\frac{\delta S_{cl}}{\delta h_{\mu \nu}} \mathcal{L}_\xi h_{\mu \nu} + \frac{\delta S_{cl}}{\delta v_\mu} \mathcal{L}_\xi v_\mu \right) = 0 \ .
\end{align}
Next, one identifies the canonical momenta $\Theta^{\mu \nu}$ and $J^{\mu}$ as in (\ref{TJ}).
Expanding out the Lie derivatives, one gets
\begin{align}
D_\mu (\Theta^{\mu\nu} + J^\mu v^\nu) = - J^\mu D^\nu v_\mu \ , \label{constrbdy}
\end{align}
where $D_\mu$ is the intrinsic covariant derivative on the slice.
In general the momentum constraints do not need to be the divergence of a symmetric tensor.
We identify the object in parentheses as the stress-energy tensor (\ref{stress}).

To compute the total stress-energy tensor using the gravitational variables, we focus on the boundary terms obtained by varying the Einstein-aether action (\ref{totalaction}) with respect to the metric and the aether fields. The variation of the usual Einstein-Hilbert action part
yields the GR Brown-York stress tensor
\begin{align}
\Theta^{BY}_{\mu \nu} = \frac{1}{8\pi G_{ae}} \left(h_{\mu \nu} K - K_{\mu \nu} \right) \label{thetaBY} \ ,
\end{align}
where $K_{\mu \nu}$ is the extrinsic curvature tensor. Now consider the variation of the vector part of the action $\int -\sqrt{-g} d^4 x K^{ABCD} \nabla_A v_C \nabla_B v_D$. We find the boundary term
\begin{align}
S^{vec}_{bdy} = \frac{1}{16\pi G_{ae}}\int d^3 x \sqrt{h}~ n_C \left(2 Y^{(CA)} v^{B} - Y^{AB} v^C \right) \delta g_{AB} \ ,
\end{align}
where $n_A$ is the unit norm to the surface (here of constant bulk coordinate $\rho=\rho_0$) and $h_{AB} = g_{AB}-n_A n_B$. For the contribution to the boundary stress-energy tensor, we find in our coordinates $x^A = (x^\mu,\rho)$
\begin{align}
\Theta^{vec}_{\mu \nu} = \frac{1}{8\pi G_{ae}} \left(-Y_{(\mu \nu)} v^C n_C + 2 n^C Y_{C(\mu} v_{\nu)} + 2 n^C Y_{(\mu|C|} v_{\nu)} \right) \label{thetavec} \ .
\end{align}

From the vector action we also find the boundary term associated with the variation of the aether co-vector
\begin{align}
S^{vec}_{bdy,vec} = - \frac{1}{8\pi G_{ae}}\int d^3 x  \sqrt{-h} ~ n_C Y^{CD} \delta v_D \ .
\end{align}
This yields the current
\begin{align}
J_\mu = - \frac{1}{8\pi G_{ae}} n^A Y_{A \mu} \label{jcurrent} \ .
\end{align}
Combining these results one gets the total boundary stress-energy tensor.
Note, that evaluation on the asymptotic boundary at infinity will require the addition of counterterms in general to remove divergent terms.

\subsection{Constraint equations}

Consider the constraint equations projected on generic surfaces of constant $\rho=\rho_0$. In GR these take the form
\begin{align}
C_\mu = G_{\mu B} n^B = 0 \ .
\end{align}
In the fluid/gravity correspondence framework, one considers solutions constructed order by order in derivatives with respect to $x^\mu$. Assuming that all the field equations are imposed at $(n-1)$ order, the $n$th order Bianchi identity 
\begin{align}
\nabla^{A}_{(0)} G^{(n)}_{A\mu} = 0 \ , 
\end{align}
can be written as a partial differential equation for the constraints as a function of radial direction $C^{(n)}_\mu(\rho, x^\mu)$. The solution 
of the differential equation is \cite{Eling:2011ct}.
\begin{align}
C^{(n)}_\mu(\rho, x^\mu) = \frac{F_\mu (x)}{A(\rho)}  \ ,
\end{align}
where $F_\mu (x)$ and $A(\rho)$ are some functions. This off-shell analysis implies the constraint equations $C_\mu=0$ have the same form on any constant $\rho=\rho_0$ slice.  At the AdS boundary, the momentum constraints are equivalent to the fluid equations $\partial^\mu T_{\mu \nu} = 0$ \cite{Bhattacharyya:2008jc}, i.e. $F_\nu(x) \sim  \partial^\mu T_{\mu \nu}$. Thus, the factorization of the field theory and radial dependence means the constraint equations projected onto any radial surface yield the same hydrodynamics equations, with identical transport coefficients.

In Einstein-aether theory the generalized Bianchi identity takes the form
\begin{align}
\nabla^A (G_{AB} - T^{ae}_{AB} + v_A E_B) + E_A \nabla_B v^A = 0 \label{genBianchi} \ ,
\end{align}
which implies the constraint equations are \cite{Jacobson:2011cc}
\begin{align}
C_\mu = (G_{A\mu} - T^{ae}_{A\mu} + v_A E_\mu) n^A = 0 \ .
\end{align}
 In this case the identity is no longer a simple conservation law, and the first term is a divergence of a non-symmetric tensor due to the $v_A E_B$ piece. Nevertheless, repeating the same analysis as for GR outlined above, shows that the constraints factorize and one can study hydrodynamics by working on any radial slice.

Instead of considering the Einstein-aether constraint equations on the universal horizon,
one can work at the Killing horizon. In this case the entropy balance law for the fluid
%
$u_\mu \partial_\nu T^{\mu \nu} = 0$,
%
can be expressed in terms of horizon variables using the null Raychaudhuri equation. The  hydrodynamic entropy balance law is equivalent to
\begin{align}
(G_{AB} - T^{ae}_{AB} + v_A E_B)\ell^A \ell^B = 0 \label{horcontract}\ ,
\end{align}
where $\ell^A$ the null normal to the horizon. For the first term we use $g_{AB} \ell^A \ell^B = 0$ and the identity
\begin{align}
G_{AB} \ell^A \ell^B = R_{AB} \ell^A \ell^B = \kappa \theta - \sigma_{AB} \sigma^{AB}  - \frac{1}{2} \theta^2 \ ,
\end{align}
where $\kappa$ the surface gravity at the Killing horizon, and $\theta$ and $\sigma_{AB}$ are the horizon expansion and shear tensor respectively.

Now it remains to evaluate the aether contributions using the field equations above. We concentrate on the contraction $(v_A \ell^A) E_B \ell^B$. Using the aether field equation (\ref{aefieldeqn}) and the form of the stress tensor, we see that the $(v_A \ell^A) \ell^C \nabla_B Y^B{}_C$ and $\lambda (v_A \ell^A)^2$ terms cancel out. The remaining pieces are
\begin{eqnarray}
(-T^{ae}_{AB} + v_A E_B)\ell^A \ell^B = c_4 (v_A \ell^A) (\ell^C \nabla_C v^B) a_B - c_4 (a_A \ell^A)^2 - \nonumber \\ \ell^A \ell^B (Y^C{}_A \nabla_C v_B - (\nabla_C v_A) Y_B{}^C - v_A \nabla_C Y_B{}^C + (\nabla_C v^C) Y_{AB} + v^C \nabla_C Y_{AB})  \ .
\end{eqnarray}

\section{Black brane hydrodynamics with $z=1$}

In this section we will analyze the hydrodynamics of the black brane solution found in
\cite{Janiszewski:2014iaa} when $z=1$, that is  $c_4=0$.
The case $z=1$ is special since the Lifshitz scaling symmetry of the boundary field theory is the same as that of relativistic CFTs, and implies the
tracefree condition on the stress-energy  tensor
\begin{align}
T^\mu_\mu = 0 \ .
\end{align}
Boost invariance, however, is still expected to be broken in the boundary field theory.
The gravitational solution is asymptotically AdS, but has a preferred time foliation and a universal horizon in the bulk interior.

In  \cite{Blas:2009yd}, it has been shown that in Horava-Lifshitz gravity the linearized spin-0 scalar perturbations around stationary background solutions generically do not propagate
 when $c_4=0$ (denoted by  $\tilde{\alpha}=0$ in (\ref{HLaction})). Moreover, as we discussed in the previous section, the divergence of the fluid  entropy current is equivalent to (\ref{horcontract}), which measures the flux of matter-energy across the horizon. The spin-0 flux is proportional to the energy density in the spin-0 waves times their speed $s_0$. The spin-0 energy density scales like $c_4$ \cite{energydensity}, while $s_0$ (\ref{speeds}) goes like $c_4^{-1/2}$. Thus the spin-0 flux scales like $\sqrt{c_4}$ and must vanish when $z=1$. In the field theory language we expect this to translate into the statement that $\alpha=0$ in this case.

  In the following
  we will study the first order hydrodynamics of this solution and show that this is indeed the case. We will calculate
  the ratio of the shear viscosity to entropy density and show that it deviates from that of Einstein gravity.
 Note, that while the first order hydrodynamics of the $z=1$ solution is a CFT hydrodynamics, this does not necessarily
 imply that a non-relativistic behavior cannot be seen in the boundary field theory beyond the hydrodynamic regime.

\subsection{Equilibrium solution}

We return to the metric and aether ansatz in (\ref{ansatz}). The $c_4=0$ solution is
\begin{align}
F(\rho) =& -\rho^2 + \frac{2\rho_h^3}{\rho} + \frac{c_3 \rho_h^6}{(1-c_3)\rho^4} \\
G(\rho) =& 1 \\
K(\rho) =& \frac{\rho^2}{(\frac{1}{\sqrt{1-c_3}}-1)\rho_h^3+\rho^3}.
\end{align}
The parameter $\rho_h$ is the value of the universal horizon. This is defined as the value where the dot product of the timelike Killing vector and the aether vanishes $\chi^A v_A = 0$. The above solution was obtained by demanding regularity at this point- that both $(\chi^A v_A)^2$ and its first derivative vanish there. The solution does not depend on $c_2$ since the covariant divergence of the aether $\nabla_A v^A=0$. This condition means all terms proportional to $c_2$ in the field equations vanish.

It has been argued \cite{Janiszewski:2014iaa,Berglund:2012fk} that there is a Hawking temperature associated with this surface
\begin{align}
T = \frac{3 \rho_h}{2\pi \sqrt{1-c_3}}. \label{universalT}
\end{align}
To determine the other thermodynamical variables, one can evaluate the boundary stress tensor found in Section \ref{boundarysection} on this solution in the limit as $\rho \rightarrow \infty$. The metric at the AdS boundary is conformal to the flat metric, i.e. $h_{\mu \nu}  = \rho_0^2 \eta_{\mu \nu}$, so one must normalize the expression by the overall conformal weight factor (in four dimensions)
\begin{align}
T^{bdy}_{\mu \nu} = \lim_{\rho_0 \rightarrow \infty} \rho_0 T^{tot}_{\mu \nu}. \label{bdystress}
\end{align}
$T^{bdy}_{\mu \nu}$ also contains divergent terms in the limit that must be subtracted off by the addition of appropriate counterterms to $T^{tot}_{\mu \nu}$. Computing the stress tensor using (\ref{thetaBY}), (\ref{thetavec}), and (\ref{jcurrent}) we find that the only counterterm needed to produce a finite answer is just the GR one,
\begin{align}
T^{counter}_{\mu \nu} = -\frac{1}{4\pi G_{ae}} h_{\mu \nu}, \label{counterstress}
\end{align}
independent of $c_3$ and proportional to the boundary metric.

 The final result is a conformal perfect fluid stress tensor
\begin{align}
T^{bdy}_{\mu \nu} =   p \left(\eta_{\mu \nu} + 3 u_\mu u_\nu \right)
\end{align}
with energy density
\begin{align}
\epsilon = 2p = \frac{\rho_h^3}{4\pi G_{ae}}. \label{epsilon}
\end{align}
This agrees with the value found by the on-shell Hamiltonian analysis in \cite{Janiszewski:2014iaa}. The value of thermal entropy density can be derived for example from the thermodynamic identity $\epsilon+p = sT$, giving
\begin{align}
s =  \frac{\rho_h^2 \sqrt{1-c_3}}{4} \label{sdensity}.
\end{align}
Note, that in the weak field regime the effective Newton constant $G_{ae}=G_N$ when $z=1$, $c_4=0$. We will set this constant to be unity.

To probe the nature of the dual system, one can study perturbations of this solution in the fluid-gravity setting. One considers the equilibrium metric and aether in a generally boosted frame
\begin{align}
ds^2 =& F(\rho) u_\mu u_\nu dx^\mu dx^\nu - 2 u_\mu dx^\mu d\rho + \rho^2 P_{\mu \nu} dx^\mu dx^\nu \\
v_A dx^A =& K(\rho) d\rho + \frac{1-F(\rho)K(\rho)^2}{2K(\rho)} u_\mu dx^\mu \ ,
\end{align}
where $u^\mu$ is the usual boost (fluid) velocity. Note that in GR, the boost we have implemented is a general coordinate transformation, which is a symmetry of the theory. However, in the Horava-Lifshitz case the symmetry group is reduced to that of coordinate transformations that preserve the preferred time foliation. Since a boost naturally changes the notion of simultaneity, the solution in the boosted frame is not physically equivalent to the one in the rest frame. 

The procedure is to then allow the velocity and Hawking temperature to be functions of the field theory coordinates: $u^\mu(x^\mu)$ and $\rho_h(x^\mu)$. Since the metric and aether are no longer solutions, one must solve the field equations order by order in derivatives of $x^\mu = (t,x^i)$ subject to asymptotically AdS and regularity in the bulk interior.

\subsection{The first order corrections}

We make the ansatz that the solution to the metric and aether field at first order in derivatives has the following form
\begin{align}
g_{AB} dx^A dx^B = F(\rho) u_\mu u_\nu dx^\mu dx^\nu - 2 u_\mu dx^\mu d\rho + \rho^2 P_{\mu \nu} dx^\mu dx^\nu \nonumber \\ - 2 J(\rho) u_\mu a_\nu dx^\mu dx^\nu + L(\rho) u_\mu u_\nu (\partial_\lambda u^\lambda) dx^\mu dx^\nu + H(\rho) \sigma_{\mu \nu} dx^\mu dx^\nu \ ,
\end{align}
while the aether is
\begin{align}
v_A dx^A =& K(\rho) d\rho + \frac{1-F(\rho)K(\rho)^2}{2K(\rho)} u_\mu dx^\mu + M(\rho) a_\mu dx^\mu - (1/2) K(\rho) L(\rho) u_\mu (\partial_\lambda u^\lambda) dx^\mu \ .
\end{align}
This assumes the standard fluid-gravity gauge choice that at all orders $g_{\rho \rho} = 0$ and $g_{\rho \mu} = -u_\mu$.  Note that this form of the first order correction is consistent with the unit vector condition on the aether field $v_A v^A = -1$. This is the most general ansatz for the metric and aether one can write down using first order hydrodynamical variables (derivatives of temperature have been traded for derivatives of $u_\mu$ using the zeroth order equations).

Using xAct \cite{xAct} one can compute the full set of Einstein-aether field equations to first order in derivatives in order to solve for the unknown functions $J(\rho)$, $L(\rho)$, $M(\rho)$,  and $H(\rho)$. We will first concentrate the solutions to $J(\rho)$, $L(\rho)$, and $M(\rho)$. The solution for $L(\rho)$ can be found from the $F_{\rho \rho}=0$, $E_\rho=0$, $u^\mu F_{\rho \mu}=0$ where $F_{AB} = G_{AB} - 3 g_{AB} + T^{ae}_{AB}$. These are a complicated set of ordinary differential equations. Solving in for example Maple, one finds the only solution is $L(\rho)= \rho$.

The field equations components $P^\nu_\mu E_\nu=0$ and $P^\nu_\mu F_{\rho \nu} =0$,  are a very complicated coupled system of differential equations for $M$ and $J$. In this case finding a simple solution is more difficult. If we impose boundary conditions for the problem, e.g. $J(\rho) \sim \rho + const. + 1/\rho + \cdots$ the solution is just $J(\rho)= \rho$ and $M(\rho) = K(\rho)\rho$. As a result, the metric and aether turn out to be conformally covariant, following \cite{oai:arXiv.org:0809.4272}. The combination
\begin{align}
A^{(1)}_\mu = a_\mu - \frac{1}{2} u_\mu (\partial_\lambda u^\lambda)
\end{align}
transforms like a connection under conformal transformations of the boundary metric. Explicitly, $g_{\mu \nu} \rightarrow e^{2\phi(x)} g_{\mu \nu}$ implies $A_\mu \rightarrow A_\mu + \partial_\mu \phi$. Under the corresponding transformation of the radial coordinate $\rho \rightarrow e^{-\phi(x)} \rho$, the combination
\begin{align}
d\rho + \rho A_\mu dx^\mu
\end{align}
transforms covariantly.

The presence of conformal symmetry in the solution is another hint that in this special case the transport coefficient $\alpha=0$. Since $\alpha$ is tied to the antisymmetric part of the hydrodynamic stress tensor, we consider the antisymmetric part of the boundary stress. In general this has the form
\begin{align}
T^{bdy}_{[\mu \nu]} = \frac{1}{8\pi G_{ae}} \left(c_3 n_A \nabla_{[\mu} v^A v_{\nu]} - c_4 (n_A v^A) a^{(v)}_{[\mu} v_{\nu]} \right)
\end{align}
Notice, that the second term proportional to $c_4$ is highly reminiscent of the $a_{[\mu} u_{\nu]}$ term in the fluid stress, but a clear matching would require a specific solution. In our case, when $c_4=0$, we found using the first order solution that the antisymmetric part makes no contribution at the boundary. The first order corrections to the stress tensor will only involve shear terms and depend on the function $H(\rho)$.

The result for the shear viscosity depends on the solution for the function $H(\rho)$. To find $H(\rho)$ we consider the following field equation $P^\sigma_\mu P^\lambda_\nu F_{\sigma \lambda} = 0$. One again we have a very complicated ordinary differential equation which we will not display here. In the limit as $c_3 \rightarrow 0$, this equation reduces to
\begin{align}
-\frac{1}{2} \rho^{-3} \left(-2\rho^2 \rho_h^3 H''-8 \rho_h^3 H + 4 \rho^4 + 6 H' \rho \rho_h^3 + \rho^5 H''- 2H \rho^3 \right) = 0.
\end{align}
The solution is
\begin{eqnarray}
H(\rho) &=& -\frac{1}{6 \rho_h} \left(-18 A \rho_h \ln(\rho)+6 A\rho_h \ln(-2 \rho_h^3+\rho^3)-6B \rho_h+ 2^{5/3} \ln(\rho-2^{1/3} \rho_h) \right.  \nonumber \\&& \left. -2^{2/3} \ln(\rho^2+ 2^{1/3} \rho_h \rho +2^{2/3}\rho_h^2)+2^{5/3} \sqrt{3} \arctan((1/3) \sqrt{3}(2^{2/3} \rho+\rho_h)/\rho_h)\right) \label{HGR}
\end{eqnarray}
Fixing AdS boundary conditions requires $B = \sqrt{3} \cdot 2^{2/3} \pi/6 \rho_h$. The asymptotic solution near the boundary at $x=1/\rho=0$ is then
\begin{align}
H(x) = 2/x + 2 \rho_h^3 A x + \rho_h^3 x^2 + 2 \rho_h^6 A x^4 + \cdots
\end{align}
Similarly, requiring that there is no divergence (curvature singularity) at the horizon fixes $A = -2^{2/3}/3\rho_h$. This matches the solution found in the literature, e.g. \cite{VanRaamsdonk:2008fp}.

For finite $c_3$ our strategy is to make the ansatz
\begin{align}
H(x) = H_0/x + H_1 + H_2 x + H_3 x^2 + \cdots \label{Hexpansion}
\end{align}
and solve for the coefficients $H_i$. The result is that
\begin{align}
H(x)= 2/x + H_2 x + \frac{1}{2} \frac{\rho_h^3 (2\sqrt{1-c_3}-c_3)}{\sqrt{1-c_3}} x^2 + \rho_h^3 H_2 x^4 + \cdots
\end{align}
Up to order $x^7$ the solution is characterized by one free parameter $H_2$. As in the pure GR case, we expect that one can tune the solution to a particular value of $H_2$ to avoid a curvature singularity at the (Killing) horizon.

For small (positive) values of $c_3$ one can work with the differential equation for $H$ to first (linear) order in $c_3$ corrections. The resulting equation can be solved analytically
\begin{eqnarray}
H(\rho) &=& H_{GR} + \frac{1}{12 \rho_h} \left(2 c_3 2^{2/3} \rho \ln(\rho-2^{1/3} \rho_h) - c_3 2^{2/3} \ln(\rho^2 + 2^{1/3} \rho \rho_h + 2^{2/3} \rho_h^2)  \right.  \nonumber \\&& \left. + 2 c_3 \sqrt{3} \cdot 2^{2/3} \arctan((1/3) \sqrt{3}(2^{2/3} \rho+\rho_h)/\rho_h) + 12 c_3 \rho_h \right) \ ,
\end{eqnarray}
where $H_{GR}$ is given by (\ref{HGR}). As a result, the value of the coefficient $H_2$ is
\begin{align}
H_2 =& \frac{2^{5/3} \rho_h^2}{3} (c_3/2-1). \label{H2}
\end{align}

\subsection{Raychaudhuri equation and the entropy law}

With a first order solution in hand, we now study the form of the Raychaudhuri equation for the Killing horizon. Since $c_4=0$,
\begin{align}
Y_{AB} = c_2 g_{AB} \nabla_C v^C + c_3 \nabla_B v_A.
\end{align}
We find that
\begin{eqnarray}
\kappa \theta - \ell^A \nabla_A \theta - \sigma_{AB} \sigma^{AB} - \frac{1}{2} \theta^2 \nonumber \\  +c_2 (\ell^B v_B) \ell^A \nabla_A (\nabla_C v^C) - c_3 (\ell^A \nabla_A v^C)(\ell^B \nabla_C v_B) + c_3 (\ell^A \nabla^C v_A)(\ell^B \nabla_C v_B) \nonumber \\ - c_3 (\nabla_C v^C) \ell^A \ell^B \nabla_A v_B + c_3 (\ell^C v_C) \ell^B \nabla_A \nabla^A v_B - c_3 \ell^A \ell^B v^C \nabla_C \nabla_B v_A = 0 \label{Ray} \ .
\end{eqnarray}
Note that the location of the Killing horizon is
\begin{align}
\rho_{KH} = \left(\frac{1-c_3+\sqrt{1-c_3}}{1-c_3}\right)^{1/3} \rho_h \ ,
\end{align}
and the surface gravity is
\begin{align}
\kappa_{KH} = \frac{3\rho_h (1+\sqrt{1-c_3})(1-c_3)^{2/3}}{(1-c_3+\sqrt{1-c_3})^{5/3}}.
\end{align}
It is also useful to transform to a coordinate system where the horizon radius is fixed at zero: $\bar{\rho} = \rho - \rho_{KH}(x)$. Transforming to these coordinates, one finds
\begin{align}
g_{AB} dx^A dx^B = F(\bar{\rho}) u_\mu u_\nu dx^\mu dx^\nu  + 2 u_\mu dx^\mu d\bar{\rho} + (\bar{\rho}+\rho_{KH})^2 P_{\mu \nu} dx^\mu dx^\nu - \nonumber \\ 2 \bar{\rho} u_\mu A_\nu dx^\mu dx^\nu + H(\bar{\rho}) \sigma_{\mu \nu} dx^\mu dx^\nu \ ,
\end{align}
and
\begin{align}
v_A dx^A =& K(\bar{\rho}) d\bar{\rho} + \frac{1-F(\bar{\rho})K(\bar{\rho})^2}{2K(\bar{\rho})} u_\mu dx^\mu + \bar{\rho} K(\bar{\rho}) A_\mu dx^\mu \ .
\end{align}
In this coordinates, the normal to the Killing horizon is $u^\mu$ to all orders.

Now we can use this ansatz to compute the Raychaudhuri equation to second order in derivatives and then evaluate at the Killing horizon, $\bar{\rho}=0$. The first step is consider the equation at first order in derivatives of $u^\mu$ and $\rho_h$. Combining terms from both the pure gravity and aether parts of the equation, we arrive at the following result
\begin{align}
\frac{3}{\rho_h (1+\frac{1}{\sqrt{1-c_3}})^{2/3}} \partial_\mu (\rho_h^2 u^\mu) = 0.
\end{align}
Up to an overall factor, this equation matches that of the entropy conservation law
\begin{align}
\partial_\mu (s u^\mu) = 0.
\end{align}
with $s$ given by (\ref{sdensity}).

The next step is to evaluate the Raychaudhuri equation at second order in derivatives.  Here we must use the other projection of the constraint equations in the horizon limit
\begin{align}
(G_{A \mu} - T^{ae}_{A \mu} + v_A E_\mu)\ell^A = 0.
\end{align}
This is equivalent to the fluid equation $P^\alpha_\nu \partial_\mu T^{\mu \nu} = 0$,
\begin{align}
P^\nu_\mu \partial_\nu \ln \rho_h + a_\mu = 0 \ ,
\end{align}
and can be imposed in the second order expressions. The result can be put into the following form
\begin{align}
\frac{3}{\rho_h (1+\frac{1}{\sqrt{1-c_3}})^{2/3}} \partial_\mu (\rho_h^2 u^\mu) - \left(\frac{4+4\sqrt{1-c_3}-4c_3-2c_3 \sqrt{1-c_3} + c_3^2}{2 (1+\sqrt{1-c_3})^2}+ \nonumber \right. \\  \left.  \frac{c_3}{8\rho^4 K^2}(\rho^2 H'-2 \rho H)\right) \sigma^2 = 0, \label{entropylaw}
\end{align}
where the functions appearing are to be evaluated at the Killing horizon. Consistent with results at the boundary, all $a_\mu a^\mu$ terms cancel out of the final expression and only shear squared remains.

\subsection{The ratio of shear viscosity to entropy density}

We now can determine the shear viscosity by matching (\ref{entropylaw}) to the general hydrodynamic entropy balance law (\ref{hydroentropylaw})
\begin{align}
\partial_\mu (s u^\mu) = \frac{2\eta}{T} \sigma_{\mu \nu} \sigma^{\mu \nu}.
\end{align}
with $s$ given in (\ref{sdensity}) and $T$ in (\ref{universalT}). We read off that
\begin{align}
\eta  = \frac{\rho_h^2}{16\pi} (1+\frac{1}{\sqrt{1-c_3}})^{2/3} \left(\frac{4+4\sqrt{1-c_3}-4c_3-2c_3 \sqrt{1-c_3} + c_3^2}{2 (1+\sqrt{1-c_3})^2}+ \frac{c_3}{8\rho^4 K^2}(\rho^2 H'-2 \rho H)\right).
\end{align}
For $c_3 > 1$ the formula is ill-defined, but in this regime the theory is known to suffer from negative energies, unstable linearized wave modes, etc. See the review in \cite{AEtheory}. Working to linear order in $c_3$ and using the results above for $H$, we find
\begin{align}
\eta = \frac{2^{2/3}}{16\pi} \rho_h^2 \left(1-c_3/2 + O(c_3^2)\right) \label{eta} \ .
\end{align}
As a check, we also evaluated the shear viscosity from the boundary stress tensor.  In this case, one inserts the first order metric solution into (\ref{bdystress}) with counterterm (\ref{counterstress}) and reads off the shear viscosity as the coefficient of the shear term in the hydrodynamic stress tensor. The result depends on the value of $H_2$ in the asymptotic expansion (\ref{H2}) for $H(\rho)$
\begin{align}
\eta = - \frac{3}{32\pi} H_2,
\end{align}
matching with (\ref{eta}), as expected.

Dividing by the entropy density $s$ (\ref{sdensity}), we find the shear viscosity to entropy density ratio
\begin{align}
\frac{\eta}{s} = \frac{2^{2/3}}{4\pi}\left(1+O(c_3^2)\right).
\end{align}
Expanding out the solution for $H(\rho)$ to higher orders in $c_3$ and repeating the calculation indicates that the ratio is independent of $c_3$ up to fourth order. Therefore we conjecture that in general
\begin{align}
\eta = \frac{2^{2/3} \sqrt{1-c_3}}{16\pi} \rho_h^2.
\end{align}
Curiously, in the limit as $c_3 \rightarrow 0$ the shear viscosity to entropy density ratio differs from the GR value of $1/4\pi$ by a factor of $2^{2/3}$. The discrepancy arises because as $c_3 \rightarrow 0$ the universal horizon entropy density $s$ goes to $\rho_h^2/4$, while the Killing horizon entropy density expected in pure GR gives $s_{KH} = A_{KH}/4 = 2^{2/3} s$. This suggests that simply taking  $c_3 \rightarrow 0$ limit in Horava-Lifshitz gravity does
not reduce to GR. The causal boundary in GR is associated with the Killing horizon, while that of
 Horava-Lifshitz gravity with the universal horzion. Thus, one should in the limit also shift variables to $s_{KH}$.

\section{Discussion}

In the following we outline several open questions and directions.
The new hydrodynamic transport associated with the breaking of boost invariance vanishes in the particular $z=1$ solution that we studied.
It would be valuable to construct gravitational solutions with $z\neq 1$, where one expects it to be generically non-vanishing.

Superfluid Lifshitz hydrodynamics has been analyzed in
\cite{Chapman:2014hja}. It would be interesting to construct dual gravitational solutions. These are black branes
in Horava-Lifshitz gravity with hair corresponding to the condensate that breaks a $U(1)$ global symmetry.

A simple formula for the ratio of the bulk viscosity  to shear viscosity in holographic Lifshitz hydrodynamics
has been derived in
\cite{Hoyos:2013cba}. It is a generalization of \cite{Eling:2011ms}
and is based on the horizon focusing equation.
It is not clear that similar formula can be worked out for the new transport coefficient associated with broken boost invariance, but
it is worth exploring this further.

If in addition to breaking boost invariance, one also allows a breakdown of rotational symmetry then there
are new expected transports in field theory hydrodynamics. These should presumably correspond to
the spin-1 helicity mode in the gravitational description.
It would be interesting to work out this relation.

Finally, while the $z=1$ black brane hydrodynamics is conformal, it would be of interest to know whether
there are Lorentz violating aspects that it exhibits beyond the hydrodynamic limit, both in the bulk and in the
boundary field theory.

\section*{Acknowledgements}
We would like to thank Igal Arav and Adiel Meyer for valuable discussions.
This work is supported in part by the I-CORE program of Planning and Budgeting Committee (grant number 1937/12), and by the US-Israel Binational Science Foundation.

\end{document}